%Paper: cond-mat/9409005
%From: Pietrino <donatis@tsmi19.sissa.it>
%Date: Fri, 02 Sep 1994 10:30:13 +0200

% The figure will be sent via ordinary mail on request to the authors.

\magnification=1200
\hfuzz=2.3pt

\font\twelvebf = cmbx12 scaled \magstep2
\font\bigbf=cmbx10 scaled \magstep2
\font \titlefont=cmbx10 scaled\magstep1
\font\smc=cmcsc10

\def\chapter#1#2{\negthinspace
{\twelvebf #1}\ \ {\twelvebf #2}
}

\def\subchapter#1#2{\negthinspace
{\bigbf #1}\ \ {\bigbf #2}
}

\def\parag#1#2{
\centerline{\bf #1\ \ #2}
}

\baselineskip=14pt

\def\p{\par\noindent}
\def\v{\rho_0}

\def\A{\vec A}
\def\At{\aem_{\theta}}
\def\AT{\acs_{\theta}}
\def\pa{\partial}
\def\pax{\pa_x}
\def\Par{\pa_r}
\def\paz{\pa_z}

\def\ACS{\A^{CS}}
\def\acs{A^{CS}}
\def\dr{(\delta\rho)^2}
\def\DR{\delta\rho}
\def\gh{\hat g}
\def\Ix{\int dx\,}
\def\Ir{\int d^2r\,}
\def\N{\vec\nabla}
\def\NW{\vec\nabla\wedge}
\def\Aem{\A^{em}}
\def\aem{A^{em}}
\def\DM{\delta\rho_{\scriptscriptstyle M}}
\def\E{\vec E}
\def\B{\vec B}
\def\D{\vec D}
\def\BM{\B_M}
\def\ev{e^2\v}
\def\ke{{ke\over 2\pi}}

\def\gsim{ \lower .75ex \hbox{$\sim$} \llap{\raise .27ex \hbox{$>$}} }
\def\lsim{ \lower .75ex \hbox{$\sim$} \llap{\raise .27ex \hbox{$<$}} }

\null
\vskip-2truecm
{{\bf  S.I.S.S.A. 50/94/EP }}
\vskip2truecm
\vskip0.5truecm
\centerline{\titlefont  MAGNETIC SCREENING PROPERTIES}
\smallskip
\centerline{\titlefont  OF AN INCOMPRESSIBLE CHIRAL FLUID}

\vskip3truecm
\centerline{Pietro Donatis and Roberto Iengo}
\vskip 1.0truecm
\centerline{{\it International School for Advanced Studies,
                I-34014 Trieste, Italy and}}
\centerline{{\it Istituto Nazionale di Fisica
               Nucleare, INFN, Sezione di Trieste, Trieste, Italy}}

\vskip3.0truecm
\noindent{\bf Abstract}
\vskip.3truecm
\par
\noindent
We study the possible penetration of a static magnetic field in an
idealized sample of many layers supporting a two dimensional charged chiral
quantum fluid, to see whether there is a kind of Meissner effect. This is a
non standard problem since the quantum fluid is incompressible having a gap
in its spectrum. We find that the system shows an intermediate behaviour
between superconducting and non-superconducting fluids, the magnetic field
being screened or not depending on its orientation relative to the layers.
\vfill
\eject

\newcount\equanumber	\equanumber=0
\def\chapterlabel{1}

\def\eqname#1{\relax\global\advance\equanumber by 1
  \xdef#1{{\rm(\chapterlabel.\number\equanumber)}}#1}

\def\eqn{\eqno\eqname}

\newcount\refnumber	\refnumber=0
\def\refname#1{\relax\global\advance\refnumber by 1
  \xdef#1{{\rm[\number\refnumber]}}#1}
\def\ref{\refname}

\null
\goodbreak
\noindent
\def\chapterlabel{1}
\chapter{1.}{Introduction.}
\bigskip
\noindent
The study of the physical properties of non-conventional quantum fluids
appears to be interesting and promising, even if a precise and detailed
comparison with a real existing system is not attempted. In particular,
attention has been devoted to possible non-relativistic quantum fluids
living in a two-dimensional space, as a framework of models which could be
related to layered superconductors (see for instance references
\ref\LAW, \ref\BULA, \ref\CLEM\ in particular for the magnetic properties
of realistic layered superconductors). A general approach consists in
introducing the problem by means of an effective lagrangian, \`a la
Landau-Ginzburg, representing an universality class which should summarize
the relevant degrees of freedom of some underlying microscopic theory not
explicitly specified. A particularly interesting class, which is peculiar
of two space dimensions, involves somehow a Chern-Simons gauge field,
breaking parity and time reversal invariance, describing a non-relativistic
quantum fluid which makes a {\it chiral} distinction between some left- and
right-handed behaviour (references \ref\MARCH, \ref\HALP, \ref\WENW,
\ref\NOI).
\par
Here we consider a particular universality class of that kind characterized
by the fact that the spectrum of the fluid (before considering it charged
and coupled to an electromagnetic field) has a gap. This chiral quantum
fluid was presented and extensively studied in ref \NOI\ (it turned out that
it is {\it formally} similar to an effective lagrangian used to describe
the Fractional Quantum Hall effect \ref\ZHK, \ref\LEEZ, \ref\LEEF,
\ref\ZHANG. In our case however, the physical context, the interpretation
and the range of the parameters are quite different). It is of course very
interesting to study the behaviour of the fluid when it is charged and can
carry electromagnetic currents, in order to compare with the behaviour of
superconductors. Therefore here we address ourselves more specifically  to
a rather crucial issue. Due to the fact that this quantum fluid has a gap
in the spectrum and it is like an ``incompressible'' fluid, does it possess
the property of screening a magnetic field like a superconductor (Meissner
effect) does? Since there is a gap, the standard mechanism similar to the
Higgs mechanism in field theory looks impossible. There is no massless
Goldstone boson which can provide a ``mass term'' for the electromagnetic
field, thus resulting in a finite penetration length. In fact, in our case,
apart from the collective modes representing the overall uniform motion of
the fluid, there are not the linearly dispersing compressional modes, which
are found in mean field treatments of fluids related to fractional
statistics, see in particular \ref\FETT, \ref\CHEN, \ref\FETTH, \ref\WENZ,
\LEEF, (for a possible different treatment, see appendix A of \NOI\ and
\ref\IEN).
\par
One can make this question more precise for the peculiar quantum fluid we
are investigating. Indeed, the reason why there is a gap is due to the
fact that it is coupled to a Chern-Simons gauge field $\ACS$. This field is
determined by the constraint that its field strength $\NW\ACS$ is
proportional to the density fluctuation $\DR$ of the fluid (see section 2
for the formulation of the effective theory). Since the fluid is
non-relativistic there is conservation of the total number of particles,
that is $\int\DR\!=\!0$, consistently with the fact that $\NW\ACS$ is
``exact'', i.e. it is a total derivative. It is because of the CS gauge
field that the Goldstone theorem is evaded and that the $U(1)$ phase
invariance breaks down, but there is a gap in the spectrum.
\par
Next, when considering the coupling with an electromagnetic field, the sum
$(\ACS\!+\!e\Aem)$ of the two gauge fields will appear in the covariant
derivative. As observed in particular in reference \ref\WENZE, it looks
like there remains an unbroken $U(1)$ symmetry, because of a simultaneous
gauge transformation of both $\ACS$ and $\Aem$ which cancels in the sum. A
``mass'' for the electromagnetic field, and thus the Meissner effect, is
recovered in the effective theory of reference \WENZE\ by means of a term
which separately breaks the gauge invariance for $\ACS$, due to the
condensation of the ``spinon pair'' component. In our effective theory,
instead, we consider a one-component quantum fluid and therefore there is
no such a term. The dynamics is therefore quite non standard.
\par
In practice, we can put the question in the following way. The term possibly
giving a ``mass'' to the electromagnetic field, which in the standard case
would be proportional to $(e\Aem)^2\rho$ (where $\rho$ is the fluid
density), is now $(\ACS\!+\!e\Aem)^2\rho$. Therefore, there is the
possibility that no ``mass'' and no screening of the magnetic field is
obtained, if the configuration
$$
\ACS=-e\Aem
\eqn\comp
$$
is energetically favourite. (To avoid possible confusion with the
literature on the Hall effect, we remind that in references \ZHK, \LEEZ,
\LEEF, \ZHANG, the relation \comp\ implements the ``filling'' relation, the
$CS$ field being proportional to the total density and the electromagnetic
field representing the strong, fixed, space-independent magnetic field of
the Hall effect. In our case instead, the $CS$ field strength is
proportional to the density {\it fluctuation}, and the space dependence of
the magnetic field is not a priori given, but has to be determined by the
variation of the hamiltonian).
\par
We consider here the case of $\Aem$ corresponding to a static magnetic
field, and the problem of its penetration in the bulk of a sample made of a
stack of many two-dimensional layers, where the quantum fluid lies (see
figure 1{\it a-b}). In ref \NOI\ we already discussed this problem, finding
a different behaviour depending on the orientation of the magnetic field
relative to the layers. Namely if the field is orthogonal to the layers'
plane, case of figure 1{\it a}, the system behaves like a type II
superconductor with a finite penetration length. In the case of figure
1{\it b}, when the field is parallel to the layers, instead the penetration
length grows with some fractional power of the sample size. Thus in this
case there is no Meissner effect, strictly speaking, in agreement with
general considerations.
\par
In the study done in reference \NOI\ we have included in the energy
computation also the electrostatic effects, arising from the fact that
locally $\DR\!\ne\!0$. We imagine of course that the layer where the fluid
lies provides a uniform background that neutralizes the charge, so that the
fluid is globally neutral. But assuming that the background charge cannot
move, there are locally electrostatic effects where $\DR\!\ne\!0$. These
effects arise in particular in the configuration of equation \comp, due to
the fact that $\DR$ is proportional to $\NW\ACS$. The resulting additional
energy helps in disfavouring the configuration \comp.
\par
However it can well occur that, in more realistic cases, the
background-neutralizing-charge is not really fixed and that it can in in
some way compensate the local excess of electrostatic charge so that the
system remains locally neutral. Thus it is important to reconsider the
problem.
\par
Here we discuss the central issue of the screening of the magnetic field,
by assuming that there are no additional electrostatic terms in the energy.
Therefore we examine whether there are intrinsic mechanisms which would
energetically disfavour the penetration of the magnetic field.
\par
Our result is that, even in this case, the pattern remains qualitatively
the same as said above, namely we find screening, i.e. finite penetration
of the magnetic field, in the configuration of figure 1{\it a} and infinite
penetration in the configuration of figure 1{\it b}, although with some
important quantitative differences.
\par
We have summarized in section 3 the main physical reasons for these
results, while section 2 summarizes the formulation of the effective
theory. Sections 4 and 5 contain the more precise and quantitative
analysis.
\par
In conclusion, we find that a chiral quantum fluid, of the
universality class described in section 2, besides having other interesting
properties discussed in \NOI, behaves with respect to a static magnetic
field as a ``quasi-superconductor'', that is in a way which is somewhat
intermediate between superconducting and non-superconducting fluids. This
could open new prospectives in the scenario of the layered quantum fluids
and the possible realistic systems.
\nobreak
\null
\nobreak
\vskip6truecm
\nobreak
\centerline{{\it Figure 1}}
\vfill
\goodbreak
\bigskip
\bigskip
\goodbreak
\noindent
\newcount\equanumber	\equanumber=0
\def\chapterlabel{2}
\chapter{2.}{The effective theory.}
\bigskip
\noindent
The theory which we discuss in this paper is described by the following
non-relativistic effective lagrangian density in two space and one time
dimensions:
$$
{\cal L}=i\phi^*\pa_0\phi-{1\over 2m}\bigl|\D\phi\bigr|^2-
{\gh\over m}(\DR)^2
\eqn\
$$
here $\phi(\vec x,t)$ is a non relativistic complex field which plays the
r\^ole of order parameter, related to the density by
$$
\rho=|\phi|^2\ .
\eqn\
$$
We will comment on the possible values of the dimensionless constant $\gh$
in the next section (in reference \NOI\ we used $g\!=\!{\gh\over m}$).
\p
We assume, as it is proper for a non-relativistic theory, a fixed total
number of particles $N$, that is we keep {\it fixed} $N\!=\!\int d^2x\rho$.
\p
$\D$ is the covariant derivative $\D\!=\N\!-\!i\ACS$ and $\ACS$ is a
Chern-Simons gauge potential related to the matter density fluctuation by
the constraint equation:
$$
\NW\ACS={2\pi\over k}\DR \qquad
\DR=\rho-\v \qquad
\v\equiv<\rho>\ .
\eqn\cern
$$
$k$ is a dimensionless number which we consider to be of the order of few
units. As a consequence of the conservation of the number of particles we
have consistently:
$$
\int d^2x \NW\ACS=0\ .
\eqn\
$$
It is seen that the small deformations have the spectrum
$$
E(p)=\sqrt{{\cal E}^2+{1\over 4m^2}(\vec p\,^4+16\gh\v\vec p\,^2)}
\eqn\
$$
where ${\cal E}\!=\!{2\pi\over mk}\v$ is the gap. The whole spectrum,
including the vortex excitations (see also reference \ref\DON), and other
relevant properties, in particular the chiral features, have been discussed
in ref \NOI.
\p
Now we are going to study the possible properties of the system of
screening an external magnetic field. Since this is essentially a three
dimensional phenomenon we suppose to build up a multilayered bulk of many
two dimensional thin films separated by a spacing $d$ (see figure 1).
\p
We are interested in stationary situations, and therefore we look for time
independent configurations described by the hamiltonian
$$
H=\int d^3x\biggl\{
{1\over 2md}\bigl|\bigl(\N-ie\Aem-i\ACS\bigr)\phi\bigr|^2+
{1\over 2}\bigl(\B^2+\E^2\bigr)+
{\gh\over md}\dr\biggr\}\ .
\eqn\zaffa
$$
Here $\B$ and $\E$ are the magnetic and electric field, respectively.
\bigskip
\bigskip
\goodbreak
\noindent
\newcount\equanumber	\equanumber=0
\def\chapterlabel{3}
\chapter{3.}{Qualitative description and summary of the results.}
\bigskip
\noindent
Before going into computational details, let us discuss the essential
points of the magnetic field penetration problem, assuming that there is no
electrostatic field. In a $y$-independent configuration, like in figure 1,
$\acs\!=\!\acs_y(x)$ and $\aem\!=\!\aem_y(x)$ being the only non-vanishing
components of the gauge fields, the phase of $\phi$ can be taken to be a
constant, conventionally zero, otherwise it would contribute an additional
positive energy (of course it will play a r\^ole instead in the vortex
configuration of section 4.2). Let us look first at the linearized form of
the equations coming from the variation of the hamiltonian (dropping higher
derivatives terms which are irrelevant for the discussion of zero modes):
$$
\eqalign{
&-\pax^2\aem+{e\v\over md}(\acs+e\aem)=0\cr
&-{2\gh\over md}\left({k\over 2\pi}\right)^2\pax^2\acs+
{\v\over md}(\acs+e\aem)=0\cr
}\eqn\
$$
One sees immediately that there is in principle a zero mode, corresponding
to the configuration of equation \comp. Thus if \comp\ could be competitive
with the standard configuration it would spoil the screening of the
external magnetic field. However to understand its relevance, one has to
take properly into account the boundary conditions and to see how the
allowed modes of $\acs$ can actually implement \comp. In the following
subsections, we will analyze, in a rather qualitative way, the
configuration \comp\ in the two geometries of figure 1{\it a-b}. A more
quantitative study is developed in sections 4 and 5. Of course after
discussing whether or not the configuration of equation \comp\ is
energetically favourite throughout the whole sample, we still have to find
the optimum configuration and describe its space dependence. We will
discuss it in detail in the following sections using variational methods,
to be more general than the linearized approximation.
\bigskip
\medskip
\goodbreak
\noindent
\subchapter{3.1.}{Magnetic field orthogonal to the layers' plane.}
\bigskip
\noindent
Let us start discussing the case of figure 1{\it a}, namely that with $\B$
orthogonal to the layers' plane. We suppose that the external magnetic
field points in the direction of positive $z$-axis, and penetrates in the
bulk in the $x$ direction (see figure 1{\it a}). We further assume that the
matter distribution is uniform in the $y$ and in the $z$ directions and we
can choose the electromagnetic and Chern-Simons gauge field pointing in the
$y$ direction:
$$
\eqalign{
&\Aem=(0,\aem,0)\cr
&\ACS=(0,\acs,0)\cr
}
\eqn\gauge
$$
and, in the gauge $\N\cdot\Aem\!=\!\N\cdot\ACS\!=\!0$, depending only on
$x$. With these choices the hamiltonian \zaffa\ becomes ($L_{x,y,z}$ being
the sample's sizes in the various directions):
$$
{H\over L_y L_z}=
\Ix\biggl\{{1\over 2md}\Bigl(|\pax\phi|^2+
\bigl|e\aem+\acs\bigr|^2\rho\Bigr)+
{1\over 2}\B^2+
{1\over 2}\E^2+
{\gh\over md}\dr\biggr\}\ ,
\eqn\ham
$$
In this way we have
$$
\aem=xB \qquad\qquad \DR={k\over 2\pi}\pax\acs\ .
\eqn\roni
$$
{}From \roni, using \comp, we get
$$
\DR=-\ke B
\eqn\simon
$$
on every point inside each layer.
\p
Of course, the system is overall electrically neutral, therefore the
support over which the quantum fluid lies in each layer acts as the
neutralizing background. If we suppose that the neutralizing background
charges are fixed, so that the fluctuation of charged matter $\DR$
cannot be locally neutralized, then we have an electric field, inside each
layer, given by the equation
$$
\N\cdot\E={e\over d}\DR
\eqn\
$$
here ${\DR\over d}$ is the three dimensional matter density (we remind that
since $\DR$ is uniform in the $z$ direction, the $z$ component of $\E$ is
always zero, thus $\E$ is effectively two dimensional).
\p
Recalling equation \cern, $\NW\ACS={2\pi\over k}\DR$, we see that $\E$ and
$\ACS$ are dual two dimensional vectors
that is
$$
E_i={ke\over 2\pi d}\epsilon_{ij}\acs_j \qquad
i=x,y\ .
\eqn\
$$
This means that the electric field contribution in \ham\ can be rewritten
$$
{1\over 2}\E^2={k^2e^2\over 4\pi^2d^2}(\ACS)^2\ .
\eqn\
$$
Therefore also when \comp\ holds we recover, through this electrostatic
term, a ``mass term'' for the electromagnetic field $\sim\!\!(\Aem)^2$.
This case has been extensively studied in \NOI\ with the conclusion that
the quantum fluid behaves, in this configuration, as a type II
superconductor.
\medskip
Here we are interested in the case when the background charge structure is
not so rigid, and we allow the system to neutralize, in some way, the
fluctuation $\DR$. Thus, we analyze the behaviour of the system dropping
the $\E^2$ term in the hamiltonian \ham. We can take into account the
energy which is spent by the system for neutralizing the charge
fluctuations, while retaining the same form for the effective hamiltonian,
by considering values for the constant $\gh$ of equation \ham\ rather
larger than the ones previously considered in reference
\NOI\ $\bigl($there, considerations based on anyon mean field theory
suggested ${\gh\!=\!\pi(1-{1\over k})}\bigr)$.
\p
We begin by observing that from equation \simon\ we see that the
cancellation \comp\ cannot hold everywhere on the two dimensional space. In
fact if this were the case we would have
$$
\Ix{\delta\rho\over d}=
{k\over 2\pi d}\Ix\NW\ACS=
-{ke\over 2\pi d}\Ix B=
-{ke\over 2\pi d}L_xB
\eqn\miss
$$
which being different from zero violates the conservation of the number of
particles.
\p
Thus, there must be somewhere an additional missing density $\DM$.
Actually, it will be concentrated on the edge of the sample, otherwise $
\acs$ would have a jump and \comp\ would no longer hold afterwards:
$$
\DM
=\ke BL_x\delta(x-L_x)\ .
\eqn\
$$
Notice that
$$
\Ix{\DM\over d}={ke\over 2\pi d}L_xB
\eqn\
$$
which exactly compensate \miss: $\int (\DR\!+\!\DM)\!=\!0$.
\p
We see that $\DM$ is very large since it is proportional to $L_x$ which is
macroscopic. Therefore a very large energy comes, for instance, from the
term in the hamiltonian which is proportional to $\int(\DM)^2$. Thus, we
foresee that the configuration of equation \comp\ will be severely
energetically disfavourite, and that the quantum fluid will essentially
behave, in the configuration of figure 1{\it a}, as a standard
superconductor. This is confirmed in the detailed analysis of section 4.
(It can also be that $\DM$, so to speak, disappears because the fluid
undergoes locally a kind of  phase transition. But if the fluid is stable
this too would cost energy, and the conclusion would be the same).
\bigskip
\medskip
\goodbreak
\noindent
\subchapter{3.2.}{Magnetic field parallel to the layers' plane.}
\bigskip
\noindent
Let us now turn to the second case in which the external magnetic field is
parallel to the layers, that is points in the $x$ direction and penetrates
the bulk in the $z$ direction (see figure 1{\it b}). Here we suppose
uniformity of the matter distribution along $y$. Again we can choose the
gauge fields pointing in the $y$ direction as in \gauge. With these
assumptions the gauge electromagnetic field is
$$
\aem=-zB
\eqn\
$$
and, if \comp\ holds,
$$
\DR={k\over 2\pi}\pax\acs=0\ .
\eqn\zero
$$
Equation \zero\ holds everywhere but at the border of the sample. In fact,
we take formally the usual boundary condition that $\ACS$ is zero at
infinity (that is outside the sample) and from equations \comp\ and
\zero\ it is constant inside each layer, i.e. independent of $(x,y)$:
$$
\acs=-e\aem(z)\,\theta(x)\,\theta(L_x-x)\ .
\eqn\
$$
$\theta(x)$ being the usual step function $=\!0$ for $x\!<\!0$, $=\!1$
for $x\!>\!0$. Therefore we have:
$$
\DR=-\ke\aem(z)\bigl[\delta(x)-\delta(L_x-x)\bigr]\ .
\eqn\
$$
Notice that now not only $\DR\!=\!0$ inside each layer, but also
$\int dx{\DR\over d}\!=\!0$, as the conservation of the total number of
particles requires.
\p
Notice also that the total amount of fluid accumulated at each border of
every layer is $\int dx{\DR\over d}\!=\!\pm\ke zB$ which remains finite for
$L_x\!\to\!\infty$.
\p
We see thus that $\DR$ is not macroscopically large and therefore we expect
that its contribution to the energy will not be large.
\p
In reference \NOI\ we have analyzed this configuration including the
electrostatic energy which arises when $\DR$ is not neutralized by the
background. In this case the electrostatic energy comes from the attraction
of the two opposite charges accumulated at the boundaries. This gives a
relatively weak effect, because the two boundaries are far apart, and the
configuration of equation \comp\ remains energetically favourite.
\p
We will see that is even more so here, when we assume, like in section 3.1,
that the background neutralizes also locally $\DR$, and accordingly we
forget the electrostatic effects. In section 5. we analyze this case in
some detail and confirm that the configuration \comp\ is indeed favourite
and find a penetration length $l_z$, in the $z$ direction, to be of the
order of
$$
l_z\sim\bigl(L_x\delta\bigr)^{1/2}
\eqn\
$$
where $L_x$ is the sample size in the direction of $B$ (we assume
$L_y\gsim L_x$) and $\delta$ is the thickness of the border region where
the charge is accumulated $\bigl($we remind, from reference \NOI, that
including the electrostatic effect one finds
$l_z\!\sim\!(L_xd^2)^{1/3}\!\times$ (logarithmic corrections)$\bigr)$.
Thus, in this case it is true that the system behaves differently from an
ordinary superconductor, where $l_z$ is finite for $L_x\!\to\!\infty$.
Notice however that still ${l_z\over L_x}\!\to\!0$ for $L_x\!\to\!\infty$.
\bigskip
\bigskip
\break
\noindent
\newcount\equanumber	\equanumber=0
\def\chapterlabel{4}
\chapter{4.}{Screening of the magnetic field orthogonal to the layers.}
\bigskip
\medskip
\nobreak
\noindent
\subchapter{4.1.}{Edge penetration.}
\nobreak
\bigskip
\nobreak
\noindent
Here we take as a starting point the hamiltonian \ham\ dropping, as said,
the electrostatic term:
$$
{H\over L_y L_z}=
\Ix\biggl\{{1\over 2md}\Bigl(|\pax\phi|^2+
|e\aem+\acs|^2\rho\Bigr)+
{1\over 2}\B^2+
{\gh\over md}\dr\biggr\}\ .
\eqn\
$$
We imagine that the region where the magnetic field is different from zero
is, in absence of the sample, the interval $-s\!\le\!x\!\le\!L_x$. The
total flux of the magnetic field is fixed thus:
$$
{\Phi\over L_y}\equiv (L_x+s)\cdot B_M=
\int\limits_{-s}^{L_x}dx\,\B=
{\rm fixed}\ .
\eqn\flux
$$
Then we redefine the zero of the energy subtracting the constant
quantity $-{1\over 2}\Ix\BM^2$, so that we can rewrite the second term in
\ham\ as:
$$
{1\over 2}\Ix \bigl(\B^2-\BM^2\bigr)=
{1\over 2}\Ix \bigl(\B-\BM\bigr)^2\ .
\eqn\
$$
Therefore the new hamiltonian is;
$$
{H\over L_y L_z}=
\Ix\biggl\{{1\over 2md}\Bigl(|\pax\phi|^2+
|e\aem+\acs|^2\rho\Bigr)+
{1\over 2}\bigl(\B-\BM\bigr)^2+
{\gh\over md}\dr\biggr\}\ .
\eqn\nham
$$
Now we suppose that the sample is placed with an edge at the origin of the
$x$ coordinate and that its length in the $x$-direction is $L_x$. We
imagine that the sample is much smaller than the region where the magnetic
field is different from zero, that is $s\!\gg\!L_x$.
\p
Here for simplicity we treat the penetration of the magnetic field as if it
were uniform, rather than exponentially decaying, and we call $l_x$ the
penetration length. Since we have fixed the total value of the magnetic
flux we have:
$$
{\Phi\over L_y}=(s+l_x)B=(s+L_x)B_M\ .
\eqn\
$$
This leads to:
$$
B={s+L_x\over s+l_x}B_M\simeq B_M\ ,
\eqn\
$$
and to:
$$
B-B_M=\biggl({s+L_x\over s+l_x}B_M-B_M\biggr)={L_x-l_x\over s+l_x}B_M\ .
\eqn\
$$
That is:
$$
{1\over 2}\int\limits_{-s}^{L_x}dx\, \bigl(\B-\BM\bigr)^2\simeq
{1\over 2}B_M^2 (L_x-l_x)\ .
\eqn\pietro
$$
\bigskip
\noindent
To check the meaningfulness of what we are doing, let us consider the case
of the standard superconductor ($\ACS\!=\!0, \rho\!=\!\v$) and see what
happens to the screening.
\p
The hamiltonian becomes:
$$
{H\over L_y L_z}=
{1\over 2}\int\limits_{-s}^{L_x}dx\,\left(\B-\BM\right)^2+
{\ev\over 2md}\int\limits_0^{l_x}dx\,\left(\aem\right)^2\ .
\eqn\elena
$$
Notice that since $\aem\!=\!xB$, we have for the second term in \elena:
$$
{\ev\over 2md}\int\limits_0^{l_x} dx\, B^2 x^2\simeq {\ev\over 2md}B_M^2
{l_x^3\over 3}
\eqn\
$$
that is, using \pietro:
$$
{H\over L_y L_z}\simeq
{1\over 2} B_M^2\biggl(L_x-l_x+{\ev\over md}{l_x^3\over 3}\biggr)\ .
\eqn\this
$$
Minimizing \this\ with respect to the penetration length $l_x$ we get the
standard value
$$
l_x=\sqrt{md\over\ev}\ .
\eqn\lstan
$$
$\v / d$ being the three dimensional mean density. So our
assumptions make sense.
\p
Substituting back in equation \this\ we get the value of the energy of the
standard configuration:
$$
{H\over L_yL_z}={1\over 2}B_M^2\biggl(L_x-{2\over 3}l_x\biggr)\ .
\eqn\relli
$$
\bigskip
\medskip
\goodbreak
\noindent
\parag{4.1.1.}{Detailed discussion of the configuration $\ACS\!=\!-e\Aem$.}
\medskip
\noindent
Coming back to the study of our non-standard quantum fluid, as we said
there is a possibility of ruining the screening of the magnetic field by
means of the cancellation \comp. If this happens we have:
$$
\Ix{\delta\rho\over d}=
{k\over 2\pi d}\Ix\NW\ACS=
-{ke\over 2\pi d}\Ix B=
-{ke\over 2\pi d}L_xB_M
\eqn\
$$
which is far from being zero.
\p
Therefore in order to have the conservation of the number of the particles
of the fluid, somewhere there must be some {\it missing fluid density}, that
we will indicate as $\DM$, accounting for the mismatch.
\p
Due to the conservation of the number of particles, it follows that
$\Ix\DR\!=\!0$, thus it is not possible that \comp\ holds strictly,
otherwise $\Ix B\!\propto\!\Ix\DR\!\ne\!0$. So we consider a configuration
where \comp\ holds as much as it is possible that is everywhere but in a
small region, say, of thickness $\delta$. Thus we suppose that $\DM$ is
concentrated in a small, microscopic, region $\delta$ around $x\!=\!x_0$,
(as we said in section 3.1, $x_0\!\sim\!L_x$), that is:
$$
\DM={ke\over 2\pi}B_ML_x{1\over\delta\sqrt{\pi}}e^{-{(x-x_0)^2\over\delta^2}}
\eqn\
$$
in such a way that
$$
\Ix\DM={ke\over 2\pi}B_ML_x\ .
\eqn\perc
$$
Let us estimate the various contribution to the energy in this
configuration.
\p
First:
$$
\Ix {1\over 2md}\bigl|\pax\phi\bigr|^2\simeq
{1\over 2md}\Ix{(\pax\DM)^2\over 4\DM}=
{1\over 4md}\,\ke B_M{L_x\over\delta^2}\ ,
\eqn\
$$
where $\phi\!=\!\sqrt{\v+\DR}$ and thus
$\pax\phi\!=\!{1\over 2}{\pax\DM\over\sqrt{\v+\DR}}\!\simeq
\!{1\over 2}{\pax\DM\over\sqrt{\DM}}$.
\p
Second:
$$
{\gh\over md}\int_0^{L_x}dx\,\biggl(-\ke B_M+\DM\biggr)^2=
{\gh\over md}\biggl(\ke\biggr)^2B_M^2L_x
\left({1\over\sqrt{2\pi}}{L_x\over\delta}-1\right)
\ .
\eqn\nesti
$$
We get the total energy
$$
{H\over L_yL_z}={1\over 4md}\,\ke B_M{L_x\over\delta^2}+
{\gh\over md}\biggl(\ke\biggr)^2B_M^2L_x
\left({1\over\sqrt{2\pi}}{L_x\over\delta}-1\right)\ .
\eqn\
$$
We see that the energy gets a contribution proportional to
$B_M^2{L_x^2\over\delta}$. Therefore compared to the energy of the standard
configuration, equation \relli, we see that the configuration implementing
the cancellation as in equation \comp\ has an energy which is larger by a
macroscopic factor.
\p
Thus, we can disregard the possibility that the configuration \comp\ holds
true throughout the whole sample.
\bigskip
\medskip
\goodbreak
\noindent
\parag{4.1.2.}{A variational analysis.}
\medskip
\noindent
We study now in some detail the penetration of the magnetic field with a
variational approach of the full hamiltonian \nham. This will allow us to
go beyond the linearized approximation, and to take into account possible
important non-linear effects. Let us make the following ansatz:
$$
\eqalign{
&B=B_0 e^{-\lambda x} \qquad \Rightarrow \qquad
\aem =-{B_0\over\lambda}e^{-\lambda x}\cr
&B_M=B_0\cr
&\acs={2\pi\over k}\int_0^x dx'\DR(x')=
-\alpha {2\pi\over k}\v x e^{-\mu x}\cr
&\DR={k\over 2\pi}\pax\acs=-\alpha\v(1-\mu x)e^{-\mu x}\cr
&\rho=\DR+\v=\v\Bigl[1-\alpha (1-\mu x)e^{-\mu x}\Bigr]\ .\cr
}
\eqn\ansatz
$$
Note that $\Ix\DR\!=\!0$.
The r\^ole of the parameter $\alpha$ is to leave free the value of $\DR$ at the
edge. The goodness of this ansatz can be checked directly on the standard
case \elena\ from which we get back the correct value \lstan.
\p
The numerical analysis indicates that $\mu\!\gg\!\lambda$. Let us assume it
for displaying a somewhat simplified expression, verifying
{\it a posteriori} that $\mu\!\gg\!\lambda$ is indeed realized.
One gets to the following expression:
$$
{md\over L_yL_z}H=
{2\pi eB_0\rho_0^2\over k\mu^2}{1\over\lambda}\alpha+
\left({\pi^2 \rho_0^3\over 2k^2\mu^3}+
{1\over 4\mu}\gh \rho_0^2+
\mu\v c(\alpha)\right)\alpha^2+
{\ev B_0^2\over 4}{1\over\lambda^3}-
{3mdB_0^2\over 4}{1\over\lambda}\ .
\eqn\sciama
$$
Here
$$
c(\alpha)={1\over 8}\int\limits_0^{\infty}dx\,
{(2-x)^2e^{-2x}\over 1-\alpha(1-x)e^{-x}}
\eqn\
$$
is a slowly varying function of $\alpha$, which for $\alpha$ small tends to
a number of order of unit. We will consider it as a constant. Minimizing
this expression with respect to $\alpha$ we find
$$
\alpha=-{1\over\lambda}QB_0\ ,
\eqn\alfa
$$
where $Q\!=\!{4\pi ek\mu\v\over 2\pi^2\rho_0^2+\gh k^2\mu^2\v+4k^2\mu^4c}$.
Substituting back in \sciama\ and minimizing now with respect to $\lambda$
we find:
$$
l_x\equiv {1\over\lambda}=\sqrt{md\over\ev}
\sqrt{1+{16\pi^2\rho_0^3\over 9mdk^2\mu^4}Q^2}+
{4\pi^2\v\over 3ek\mu^2}Q\ .
\eqn\vert
$$
We see from equation \vert\ that the penetration length approaches the
standard one for large $\gh$, for which $Q\!\to\!0$. We recall again that a
large value of $\gh$ is expected because it effectively represents the fact
that the system must spend energy to remain electrically neutral when
$\DR\!\ne\!0$. \p Indeed this results have been confirmed through a
numerical minimization with respect to $\lambda$ and $\alpha$ of the
hamiltonian which is obtained from \nham\ using the ansatz \ansatz, with
$\mu\!=\!\sqrt{\rho_0}$ (corresponding to a coherence length of
$\sim 10{\rm\AA}$), for various values of $B_0$. Already for $\gh\!=\!1.5$
we get $l_x\!\simeq\!1.2\cdot10^3{\rm\AA}$, and for $\gh\!=\!10$ we get
$l_x\!\simeq\!9.8\cdot 10^2{\rm\AA}$ (the standard value equation
\lstan\ is $l_x\!\simeq\!8.5\cdot 10^2{\rm\AA}$). We have also numerically
verified that taking $\mu$ smaller increases the energy of the
configuration, confirming that $\mu\!\gg\!\lambda$ as stated above. In
fact, for $\mu\!\to\!\infty$ we see that
$Q\!\sim\!{1\over c}{\pi e\v\over k\mu^3}$ thus $\alpha\!\to\!0$, and
$l_x\!\to\!\sqrt{\ev\over md}$. We see thus from equation \sciama\ that the
value $\mu\!\to\!\infty$ formally corresponds to the minimal energy. We
have taken $\mu$ at its physically reasonable maximum value, that is
$\mu\!\sim\!\sqrt{\rho_0}$.
\p
One can check that $\alpha$ is indeed small for that value, for
$\gh\!\sim\!10$ and $B_0\lsim 10^3$ gauss one gets $\alpha\lsim 10^{-2}$
(in all these numerical computations we have taken $m$ to be the mass of
the electron, $\sim 250{\rm\AA}^{-1}$, and
$\v\!=\!4\cdot 10^{-3}{\rm\AA}^{-2}$).
\p
We see that the penetration length is independent of the value of
$B_0$ and that $\alpha$ is proportional to $-B_0$ (that is, $\ACS$ has the
sign opposite to $\Aem$, as if the system would like the configuration
\comp\ as far as it is possible).
\bigskip
\bigskip
\goodbreak
\noindent
\subchapter{4.2.}{Vortices.}
\bigskip
\noindent
In this section we study the penetration of the magnetic field from
magnetic vortices.
The starting point will be the following hamiltonian:
$$
{H\over L_z}=
\Ir\biggl\{{1\over 2md}\Bigl|\Bigl(\N-ie\Aem-i\ACS\Bigr)\phi\Bigr|^2+
{1\over 2}\B^2+
{\gh\over md}\dr\biggr\}\ .
\eqn\hvort
$$
We look for solutions of the form:
$$
\phi(r,\theta)=f(r)e^{in\theta}\ ,
\eqn\vort
$$
$n$ is integer and represent the vorticity.
\p
Substituting \vort\ in \hvort\ we get:
$$
{H\over L_z}=
\Ir\biggl\{{1\over 2md}\biggl[\bigl(\Par f)^2+
{1\over r^2}\bigl(n-er\At-r\AT\bigr)^2f^2\biggr]+
{1\over 2}\B^2+
{\gh\over md}\dr\biggr\}\ .
\eqn\simo
$$
Finiteness of \simo\ requires:
$$
\eqalign{
&f(0)=0 \qquad\Rightarrow\qquad \DR(0)=-\v\cr
&\lim_{r\to\infty}\DR=0 \qquad\Rightarrow\qquad
\lim_{r\to\infty}f=\sqrt{\rho_0}\cr
&\lim_{r\to\infty}\At={n\over er}\cr
&\lim_{r\to\infty}\AT=0\ .\cr
}
\eqn\fran
$$
We solve this hamiltonian in a variational way with the following ansatz,
which satisfies \fran:
$$
\eqalign{
&er\At =n\Bigl(1-e^{-\lambda^2 r^2}\Bigr)
\qquad\Rightarrow\qquad
B={1\over r}\Par(r\At)={2n\over e}\lambda^2 e^{-\lambda^2 r^2}\cr
&r\AT=-{\pi\over k}\v r^2 e^{-\mu^2 r^2}\cr
&\DR={k\over 2\pi}{1\over r}\Par(r\AT)=-\v(1-\mu^2 r^2)e^{-\mu^2 r^2}\cr
&f^2=\v-\v(1-\mu^2 r^2)e^{-\mu^2 r^2}\ .\cr
}
\eqn\ans
$$
Notice that $\int_0^{\infty}d^2r\,\DR\!=\!0$, therefore there is no missing
$\DM$. Later on, we will compare this configuration with a configuration
where $\ACS\!=\!-e\Aem$ and we will need $\DM\!\ne\!0$ like in the
discussion of section 4.1.1.
\p
Notice also that in this case, differently from the case treated in the
previous section the value of $\DR$ at the origin is fixed to zero by the
requirement of finite energy, see equation \simo, so there is no $\alpha$
parameter.
\p
With this ansatz we get, supposing $\mu\!\gg\!\lambda$, to be later
verified:
$$
{H\over L_z}\simeq
{\v\pi\over 2md}\biggl({6n\pi\v\over k\mu^2}+
n^2\log{\mu^2\over 2\lambda^2}+
{2mdn^2\lambda^2\over\ev}+
{\v\gh\over 2\mu^2}\biggr)\ .
\eqn\netti
$$
Minimizing \netti\ with respect to $\lambda^2$ one finds easily:
$$
\lambda^2={\ev\over 2md}\ .
\eqn\
$$
Therefore we recover the penetration length we had found in \NOI\ studying
the case including the electrostatic interaction:
$$
l_x={1\over\lambda}=\sqrt{2md\over\ev}\ .
\eqn\recov
$$
Notice that this penetration length is independent of $\gh$.
\p
Minimizing \netti\ with respect to $\mu$ we find:
$$
\mu^2={6\pi\v\over nk}+{\gh\v\over 2n^2}\ ,
\eqn\
$$
which corresponds to a coherence length:
$$
r_v={1\over\mu}=
\sqrt{2kn^2\over 12n\pi\v+k\v\gh}\ .
\eqn\cohe
$$
Notice that, as is reasonably expected, the radius $r_v$ of the vortex
decreases with increasing $\gh$.
All these results have been confirmed through a numerical minimization with
respect to $\lambda$ and $\mu$ of the exact hamiltonian with various
values of $\gh$. Notice that the results for the vortices agrees fairly
well with those for the edge penetration.
\p
In particular for $\gh\!=\!2$ we have $r_v\!\sim\!8{\rm\AA}$ and
$l_x\!\sim\!1167{\rm\AA}$, whereas for
$\gh\!=\!10$ we have $r_v\!\sim\!6{\rm\AA}$ and the same value of
$l_x$ $\bigl($corresponding to the value of equation \recov$\bigr)$.
Notice that these results confirm that $\mu\!\gg\!\lambda$ as stated
above.
\p
Substituting \recov\ and \cohe\ in \netti\ we find the energy of the
configuration of one vortex:
$$
{H\over L_z}\simeq
{\v\pi\over 2md}\Bigl({6n\pi\v\over k}r_v^2+
n^2\log{md\over\ev r_v^2}+
n^2+
{\gh\v r_v^2\over 2}\Bigr)\ .
\eqn\paola
$$
\bigskip
\medskip
\goodbreak
\noindent
\parag{4.2.1.}{The $\ACS\!=\!-e\Aem$ configuration with rotational
symmetry.}
\medskip
\noindent
We now discuss, like in section 4.1.1, the possible cancellation \comp\ in
the case of a configuration which has a rotational symmetry, and the flux
of $B$ is given, like for the vortices case discussed above. We will
compare it with the standard vortex configuration of the previous section.
Following the discussion done there, it is unavoidable a missing $\DM$. We
suppose to have a fixed value of the magnetic flux.
$$
\Phi(B)=\Ir B={2\pi \over e} N\ .
\eqn\
$$
Therefore
$$
\Ir\DR={k\over 2\pi}\Ir\NW\ACS=
-\ke\Ir B=-\ke \pi R_0^2B=-kN\ ,
\eqn\fabio
$$
where $R_0$ is the radius of the sample, supposed to be a disk.
\p
So to have conservation of the number of particles we need a $\DM$ of the
form:
$$
\DM={kN\over 2\pi R_0}{1\over \delta\sqrt{\pi}}e^{(r-R_0)^2\over\delta^2}
\ ,
\eqn\uffa
$$
where the thickness $\delta$ is supposed to be microscopic. Note that
$$
\Ir\DM=kN.
\eqn\
$$
We take $\DM$ concentrated at a macroscopic distance from the vortex core,
say at the edge of the sample, that is at $r\!=\!R_0$. Otherwise, if it were
concentrated at some different point, say at $r\!=\!r_0$, then $\ACS$ would
have a jump of ${N\over r}$ at $r\!=\!r_0$, due to the fact that
$\Par(r\acs)\!=\!{2\pi\over k}r\DR$, and would no longer cancel $e\aem$ for
$r\!>\!r_0$, contrary to our hypothesis that the cancellation holds
throughout the macroscopic size of the sample.
\p
Let us estimate the most relevant contributions to the energy ${H\over L_z}$
coming from the presence of $\DM$. We write:
$$
{H\over L_z}=
\biggl({H\over L_z}\biggr)_1+
\biggl({H\over L_z}\biggr)_2+
\biggl({H\over L_z}\biggr)_3\ .
\eqn\
$$
We find first
$$
\biggl({H\over L_z}\biggr)_1={1\over 2md}\Ir |D\phi|^2\simeq
{1\over 8md}\Ir {(\Par\DM)^2\over\DM}=
{1\over 8md}{kN\over\delta^2}\ .
\eqn\
$$
Second:
$$
\biggl({H\over L_z}\biggr)_2=
{\gh\over md}\Ir(\DR)^2=
{\gh\over md}\biggl[\sqrt{\pi}\biggl({k\over 2\pi}\biggr)^2
{N^2\over R_0\delta}-
{k^2\over\pi}{N^2\over R_0^2}\biggr]\ .
\eqn\fabr
$$
Third:
$$
\biggl({H\over L_z}\biggr)_3={1\over 2}\Ir B^2=
{2\pi^2\over e^2}{N^2\over\pi R_0^2}\ .
\eqn\izio
$$
We can distinguish two cases.
\medskip
\p
a) $N\!=\!1$ (or few units). Then the most relevant energy contribution due
to $\DM$ comes from $\bigl({H\over L_z}\bigr)_1$, since the other two
pieces are suppressed, at least, by the factor ${1\over R_0}$, with $R_0$
macroscopic. This energy (for $\delta^2\!\sim\!{1\over\v}$) is less or
equal to the free energy of the standard configuration \paola. Therefore
for a small flux, that is for a very small magnetic field, the configuration
where $\ACS$ cancels $e\Aem$ is possibly favourite.
\medskip
\p
b) Now we suppose $B$ macroscopic, in other words $B$ is fixed in the
macroscopic limit $R_0\!\to\!\infty$. Therefore from equation
\fabio\ $N\!\sim\!BR_0^2\!\to\!\infty$. In this case the configuration
where $\ACS$ cancels $e\Aem$ gets the most relevant energy from
$\bigl({H\over L_z}\bigr)_2$, namely from the piece proportional to
${N^2\over\delta R_0}$. Therefore, for this configuration
$$
{H\over L_z}\simeq{\gh\over 4md}\sqrt{\pi}\biggl(\ke\biggr)^2 B^2
{R_0^3\over\delta}+{\rm less\, important}\ .
\eqn\
$$
We have to compare it with the
energy of the standard vortex configuration, equation \paola\ multiplied by
$N$, that is:
$$
{H\over L_z}\simeq c\,{\v\pi\over 2md}BR_0^2\ ,
\eqn\
$$
where $c$ is of the order of $\v r_v^2$, i.e. a finite number. Since
${R_0\over\delta}\!\to\!\infty$, clearly the standard configuration, or
also a configuration of many standard vortices, is energetically favourite.
\bigskip
\bigskip
\goodbreak
\noindent
\def\chapterlabel{5}
\newcount\equanumber	\equanumber=0
\chapter{5.}{Weak screening of the magnetic field parallel to the layers.}
\nobreak
\bigskip
\medskip
\noindent
In this chapter we put our attention to the configuration in which the
field $\B$ is parallel to the layers' plane (see figure 1{\it b}) and study
the screening effects in absence of the electrostatic interaction, as
discussed in the previous chapter.
\p
With considerations very similar to those made at the beginning of section
4.1 we get, in this configuration, to the hamiltonian:
$$
{H\over L_y}=\int dxdz\,\biggl\{{1\over 2md}\biggl[\bigl|\pax\phi\bigr|^2+
c_{\scriptscriptstyle J}\bigl|\paz\phi\bigr|^2\biggr]+
{1\over 2md}\Bigl|e\aem+\acs\Bigr|^2\rho+
{1\over 2}(B-B_M)^2+
{\gh\over md}\dr\biggr\}\ ,
\eqn\
$$
here $c_{\scriptscriptstyle J}$ is a constant accounting for the Josephson
coupling between the layers.
\p
We study the possible screening starting from the configuration in which we
have the cancellation \comp\ inside the sample, that is:
$$
\acs(x,z)=-e\aem(z)\bigl[\theta(x)\theta(L_x-x)\bigr]\ .
\eqn\
$$
This yields (see section 3.2):
$$
\DR(x,z)=
{k\over 2\pi}\pax\acs=
-\ke\aem\bigl[\delta(x)-\delta(L_x-x)\bigr]\ .
\eqn\
$$
Here, as in chapter 4, we suppose that the fluid density is confined in a
microscopic region of thickness $\delta$ so we approximate the
$\delta$-functions with
$$
\delta(x)\simeq {1\over \sqrt{\pi}\delta}e^{-{x^2\over\delta^2}}\ .
\eqn\
$$
Therefore we have:
$$
\DR(x,z)=
-\ke{1\over \sqrt{\pi}\delta}\biggl[e^{-{x^2\over\delta^2}}-
e^{-{(L_x-x)^2\over\delta^2}}\biggr]\aem(z)\ .
\eqn\
$$
Keeping this configuration, let us assume that the magnetic field
penetrates in the $z$ direction within a length $l_z$ and let us estimate
it. We begin by estimating the various contributions to the energy in this
configuration disregarding the terms proportional to $e^{-{L_x^2\over
2\delta^2}}\!\simeq\!0$. First:
$$
{1\over 2md}\int dxdz\, \bigl|\pax\phi\bigr|^2=
{1\over 4md\v}{1\over\sqrt{2\pi}}\biggl(\ke\biggr)^2
{1\over\delta^3}\int_0^{l_z} dz\,\Bigl(\aem(z)\Bigr)^2\ ,
\eqn\
$$
where we used the fact that
$\pax\phi\!=\!{1\over 2}{\pax(\DR)\over\sqrt{\DR+\v}}\!\simeq\!
{1\over 2\sqrt{\rho_0}}\pax(\DR)$.
\p
Second:
$$
{c_{\scriptscriptstyle J}\over 2md}\int dxdz\,\bigl|\paz\phi\bigr|^2=
{c_{\scriptscriptstyle J}\over 8md\v}
{1\over\sqrt{2\pi}}\biggl(\ke\biggr)^2{1\over\delta}
\int_0^{l_z} dz\,\Bigl(\paz\aem(z)\Bigr)^2\ .
\eqn\
$$
Third:
$$
{\gh\over md}\int dxdz\,\dr=
{\gh\over md}{1\over\sqrt{2\pi}}\biggl(\ke\biggr)^2{1\over\delta}
\int_0^{l_z} dz\,\Bigl(\aem(z)\Bigr)^2\ .
\eqn\
$$
Fourth:
$$
{1\over 2}\int dxdz\, (B-B_M)^2=
{1\over 2}B_M^2L_x(L_z-l_z)\ ,
\eqn\chiara
$$
$\bigl($here to get \chiara\ we have used arguments similar to those that
led to \pietro$\bigr)$.
\p
Now putting it all together, and using the fact that $\aem(z)\!=\!-zB_M$,
we arrive at:
$$
{H\over L_y}=
{1\over md}{1\over\sqrt{2\pi}}\biggl(\ke\biggr)^2{1\over\delta}
\biggl({1\over 4\v\delta^2}+\gh\biggr)B_M^2{l^3_z\over 3}+
{c_{\scriptscriptstyle J}\over 8md\v}{1\over\sqrt{2\pi}}
\biggl(\ke\biggr)^2{1\over\delta}B_M^2l_z+
$$
$$
+{1\over 2}B_M^2L_x(L_z-l_z)\ .
\eqn\coca
$$
Minimizing equation \coca\ with respect to $l_z$ we find:
$$
l_z\simeq\sqrt{L_x\over\Omega}\ ,
\eqn\
$$
where $\Omega\!=\!{2\over md}{1\over\sqrt{2\pi}}\bigl(\ke\bigr)^2
{1\over\delta}\bigl({1\over 4\v\delta^2}+\gh\bigr)$.
\p
This means that when the magnetic field is parallel we do not have
screening in the usual sense because $l_z\!\to\!\infty$ for
$L_x\!\to\!\infty$. But we have a ``quasi screening'' in the sense that
${l_z\over L_x}\!\to\!0$.
\p
Substituting this result back in \coca, we find the energy of the
configuration:
$$
{H\over L_y}=
{1\over 2}B_M^2L_xL_z-
{1\over 3}B_M^2(L_x)^{3/2}(\Omega)^{-1/2}\ .
\eqn\cola
$$
Let us compare this result to the energy of the standard configuration.
That is, let us take $\acs\!=\!0$ and $\phi\!=\!\sqrt{\rho_0}$; then the
hamiltonian becomes:
$$
{H\over L_y}={1\over 2}B_M^2L_x\biggl({\ev\over 3md}l_z^3+L_z-l_z\biggr)\ ,
\eqn\
$$
which is minimal for the standard value \lstan\ of the penetration length.
The value of this minimal energy is:
$$
{H\over L_y}={1\over 2}B_M^2L_x\Bigl(L_z-{2\over 3}l_z\Bigr)\ .
\eqn\sprite
$$
Comparing this equation with equation \cola\ we see that \sprite\ is
always greater than \cola\ for $L_x$ macroscopic. So the configuration
\comp, in the geometry as in fig 1{\it b} is always favourite.
\goodbreak
\bigskip
{\bf Acknowledgements.} We would like to thank Arturo Tagliacozzo for
useful comments on our previous work, which have stimulated the present
paper.
\goodbreak
\bigskip
\bigskip
\bigskip
\centerline{\bf{References.}}
\nobreak
\bigskip

\noindent
\LAW\ {\smc W.E. Lawrence, S. Doniach},
{\it in} ``Proceedings of the  XII International  Conference on Low
Temperature Physics'' (E. Kanda, Ed.), p.361, Academic Press, Tokio, 1971.
\vskip 0.3truecm

\noindent
\BULA\ {\smc L.N. Bulaevskii},
{\it Int. J. Mod. Phys.} {\bf B4} (1990), 1849.
\vskip 0.3truecm

\noindent
\CLEM\ {\smc J.R. Clem},
{\it Phys. Rev.} {\bf B43} (1991), 7837.
\vskip 0.3truecm

\noindent
\MARCH\ {\smc J. March-Russell, F. Wilczek},
{\it Phys. Rev. Lett.} {\bf 61} (1988), 2066.
\vskip 0.3truecm

\noindent
\HALP\ {\smc B.I. Halperin, J. March-Russell, F. Wilczek},
{\it Phys. Rev.} {\bf B40} (1989), 8726.
\vskip 0.3truecm

\noindent
\WENW\ {\smc X.G. Wen, F. Wilczek, A. Zee},
{\it Phys. Rev.} {\bf B39} (1989), 11413.
\vskip 0.3truecm

\noindent
\NOI\ {\smc P. Donatis, R. Iengo},
{\it Nucl. Phys.} {\bf B[FS]415} (1994), 630.
\vskip 0.3truecm

\noindent
\ZHK\ {\smc S.C. Zhang, T.H. Hansson, S. Kivelson},
{\it Phys. Rev. Lett.} {\bf 62} (1989), 82.
\vskip 0.3truecm

\noindent
\LEEZ\ {\smc D.H. Lee, S.C. Zhang},
{\it Phys. Rev. Lett.} {\bf 66} (1991), 1220.
\vskip 0.3truecm

\noindent
\LEEF\ {\smc D.H. Lee, M.P.A. Fisher},
{\it Int. J. Mod. Phys.} {\bf B5} (1991), 2675.
\vskip 0.3truecm

\noindent
\ZHANG\ {\smc S.C. Zhang},
{\it Int. J. Mod. Phys.} {\bf B6} (1992), 25.
\vskip 0.3truecm

\noindent
\FETT\ {\smc A.L. Fetter, C.B. Hanna, R.B. Laughlin},
{\it Phys. Rev.} {\bf B39} (1989), 9679.
\vskip 0.3truecm

\noindent
\CHEN\ {\smc Y.H. Chen, F. Wilczek, E.Witten, B.I. Halperin},
{\it Int. J. Mod. Phys.} {\bf B3} (1989), 1001.
\vskip 0.3truecm

\noindent
\FETTH\ {\smc A.L. Fetter, C.B. Hanna, R.B. Laughlin},
{\it Phys. Rev.} {\bf B40} (1989), 8745.
\vskip 0.3truecm

\noindent
\WENZ\ {\smc X.G. Wen, A. Zee},
{\it Phys. Rev.} {\bf B41} (1990), 240.
\vskip 0.3truecm

\noindent
\IEN\ {\smc R. Iengo, K. Lechner},
{\it Nucl. Phys.} {\bf B384} (1992), 541.
\vskip 0.3truecm

\noindent
\WENZE\ {\smc X.G. Wen, A.Zee},
{\it Phys. Rev. Lett.} {\bf 62} (1989), 2873.
\vskip 0.3truecm

\noindent
\DON\ {\smc P. Donatis, R. Iengo},
{\it Phys. Lett.} {\bf B320} (1994), 64.
\vskip 0.3truecm

\vfill
\eject
\bye